**Title**

**Learning to Reconstruct Accelerated MRI Through K-space Cold Diffusion without Noise**

**Authors/affiliations**

Guoyao Shen [1, 2], Mengyu Li [1, 2], Chad W. Farris [3], Stephan Anderson [2, 3], Xin Zhang*[1, 2]

[1] Department of Mechanical Engineering, Boston University, Boston, MA 02215.

[2] The Photonics Center, Boston University, Boston, MA 02215.

[3] Department of Radiology, Boston Medical Center and Boston University Chobanian & Avedisian School of Medicine, Boston, MA, 02118.

**Corresponding author:** Prof. Xin Zhang, email: xinz@bu.edu

**Abstract**

Deep learning-based MRI reconstruction models have achieved superior performance these days. Most recently, diffusion models have shown remarkable performance in image generation, in-painting, super-resolution, image editing and more. As a generalized diffusion model, cold diffusion further broadens the scope and considers models built around arbitrary image transformations such as blurring, down-sampling, etc. In this paper, we propose a k-space cold diffusion model that performs image degradation and restoration in k-space without the need for Gaussian noise. We provide comparisons with multiple deep learning-based MRI reconstruction models and perform tests on a well-known large open-source MRI dataset. Our results show that this novel way of performing degradation can generate high-quality reconstruction images for accelerated MRI.

**Introduction**

Recent years, deep learning-based reconstruction from under-sampled measurements for magnetic resonance imaging (MRI) has been an important topic. Although MRI is a critical tool that provides invaluable diagnostic information, long acquisition times for MRI leads to multiple limitations of this technique including delays in diagnosis, limited availability for patients, and degradation of imaging, among others. Multiple works have focused on utilizing convolutional neural network (CNN)-based models for under-sampled MRI reconstruction. Deep learning-based MRI reconstruction models have achieved superior performance these days [1, 2, 3, 4, 5, 6, 7, 8].

Most recently, diffusion models have shown greater potential and remarkable performance in various fields. Works such as denoising diffusion probabilistic models (DDPM) [9] and score-based model [10, 11] have shown promising results in image generation, in-painting, super-resolution, image editing and more [12, 13, 14, 15, 16, 17, 18, 19, 20, 21]. Diffusion models typically include two components: image degradation (forward process or diffusion process) and image generation (reverse process). In the first part, noise is gradually added to the original data in a step-by-step manner. In the second part, a deep neural network is trained as a restoration operator to perform denoising and recover the original data distribution. Recently, multiple works have been reported to utilize diffusion-based models for accelerated MRI reconstruction [22, 23, 24, 25]. Chung et al. [26] propose a diffusion model that performs accelerated MRI reconstruction from the under-sampled image directly. Xie et al. [27] propose a measurement-conditioned denoising diffusion probabilistic model (MC-DDPM) for accelerated MRI reconstruction. Instead of performing a reverse process on the image, MC-DDPM performs reconstruction on the measurement space (k-space for MRI). During the diffusion process, MC-DDPM gradually adds Gaussian noise to the k-space data conditioned on the under-sampling mask. Then, during the reverse process, MC-DDPM recovers the k-space data step-by-step.

Conventional diffusion models utilize Gaussian noise during training and generation. They can be seen as a random walk around the image density function using Langevin dynamics. Cold diffusion models further broaden the scope and consider models built around arbitrary image transformations such as blurring, down-sampling, in-painting, snowification, etc. In such way, the cold diffusion model can be considered as a generalized diffusion model which requires no Gaussian noise during training or testing [28]. The deep neural networks in cold diffusion models are trained to invert such image deformation rather than remove the Gaussian noise. The degradations in cold diffusion can be randomized or deterministic and designed as needed. This framework provides a more diverse paradigm of diffusion models beyond just the Gaussian noise and gives more flexibility for image degradation and generation.

In this work, we present a k-space cold diffusion model for accelerated MRI reconstruction, demonstrated in Figure 1. Different from previous diffusion-based models for MRI reconstruction that utilized Gaussian noise, our model performs degradation in k-space during the diffusion process. A deep neural network is trained to perform the reverse process to recover the original fully sampled image. In such a way, the k-space sampling process is integrated directly into the image degradation process, enhancing the model's generalizability, especially when the sampling process is similar. This allows for quicker application and better performance in zero-shot or few-shot learning scenarios. We provide comparisons with multiple deep learning-based MRI reconstruction models and perform tests on a well-known large open-source MRI dataset: fastMRI [29]. Our results show that this novel way of performing degradation can generate high-quality reconstruction images for accelerated MRI. We hope that this work can invoke more generalized diffusion models for MRI reconstruction in the future.

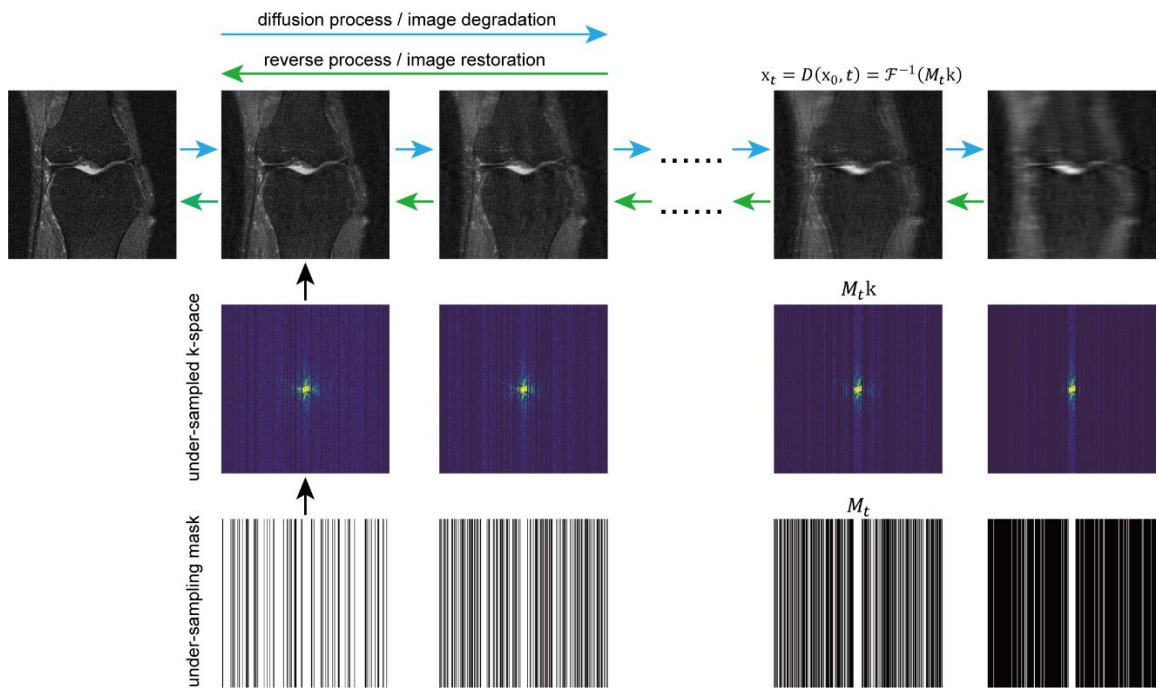

**Figure 1.** The overall pipeline of our k-space cold diffusion model.

## Methods

### *1. Cold diffusion*

Diffusion model is a class of latent variable models which uses a Markov chain to convert the noise distribution to the data distribution. It has the form $p_\theta(x_0) := \int p_\theta(x_{0:T})\, dx_{1:T}$, where $x_1, \ldots, x_T$ are latent variables of the same dimensionality as the data distribution $x_0 \sim q(x_0)$. The reverse process $p_\theta(x_{0:T})$ is a joint distribution and is defined as a Markov chain with learned Gaussian transitions starting with $p_\theta(x_T) = \mathcal{N}(x_T; \mathbf{0}, \mathbf{I})$:

$$p_\theta(x_{0:T}) := p_\theta(x_T) \prod_{t=1}^{T} p_\theta(x_{t-1}|p_\theta(x_t)), \quad p_\theta(x_{t-1}|x_t) := \mathcal{N}(x_{t-1}; \boldsymbol{\mu}_\theta(x_t, t), \sigma_t^2 \mathbf{I}) \quad (1)$$

The forward or diffusion process is the approximate posterior $q(x_{1:T}|x_0)$, which is fixed to a Markov chain that gradually adds Gaussian noise to the data according to a variance schedule $\beta_1, \ldots, \beta_T$:

$$q(x_{1:T}|x_0) := \prod_{t=1}^{T} q(x_t|x_{t-1}), \quad q(x_t|x_{t-1}) := \mathcal{N}(x_t; \sqrt{1-\beta_t}\, x_{t-1}, \beta_t \mathbf{I}) \quad (2)$$

In practice, the forward process is achieved by gradually adding Gaussian noise following the variance schedule. This process does not encounter learnable parameters. The reverse process, on the other hand, is implemented with a learnable deep neural network.

Cold diffusion model [28] is a generalized diffusion model that provides more flexibility for image degradation and restoration. Given an image data $x_0$, consider a degradation operator $D$ with severity $t$, then the degraded $x_t = D(x_0, t)$ should vary continuously in $t$. And the degradation should satisfy $D(x_0, 0) = x_0$. In the standard diffusion model, $D$ gradually adds Gaussian noise with a variance proportional to $t$. Works have been done for more efficient noise scheduling along $t$ and faster convergence [30, 31, 32, 33]. To revert this process and generate an image, the restoration operator $R$ (approximately) inverts $D$ and has the property of $R(x_t, t) \approx x_0$. The restoration operator $R$ is implemented via a deep neural network in practice and parameterized by $\theta$. This network can then be trained via the minimization problem:

$$\min_\theta \mathbb{E} \| R_\theta(D(x, t), t) - x \| \quad (3)$$

Once the degradation is chosen and the network is trained properly to perform the restoration, the network can be used to sample images from the degraded image. For standard diffusion models, one can generate images from noises as the network is trained to perform reconstruction under Gaussian noise. Standard diffusion models perform image generation (sampling from the model) by iteratively applying the denoising and adding the noise back:

$$\hat{x}_0 = R(x_t, t)$$

$$x_{t-1} = D(\hat{x}_0, t-1), \qquad t = T, T-1, \dots, 1 \tag{4}$$

The sampling strategy above works well when the restoration operator is perfect, meaning $R(D(x_0, t), t) = x_0$ holds for all $t$. However, if the restoration operator is an approximate inverse of the degradation, $x_t$ can wander away from $D(x_0, t)$ and lead to an inaccurate reconstruction. Cold diffusion proposed an improved sampling strategy. Instead of sampling $x_t$ from $\hat{x}_0$ directly, it samples $x_t$ via intermediate variables:

$$x_{t-1} = x_t - D(\hat{x}_0, t) + D(\hat{x}_0, t-1) \tag{5}$$

This sampling strategy is beneficial especially when the higher order terms in the Taylor expansion of the degradation $D(x, t)$ is non-negligible. This sampling strategy enables more reliable reconstructions for cold diffusion models with smaller total step number $T$ and a variety of image restoration operations such as deblurring, inpainting, super-resolution, snowification, etc.

### *2. Cold diffusion in k-space*

An MR scanner performs imaging by acquiring measurements in the frequency domain (i.e., k-space) using receiver coils. The relationship between the underlying image **x** and the measured k-space k can be represented as:

$$k = \mathcal{F}(x) + \epsilon \tag{6}$$

where $\mathcal{F}$ is the Fourier transformation operator and $\epsilon$ is the measurement noise. The MRI acquisition speed is limited by the amount of k-space data to obtain. This acquisition process can be accelerated by down-sampling only a portion of the k-space data $\tilde{k} = M \circ k$, where $M$ is a down-sampling binary mask that selects a subset of the overall k-space and ∘ indicates Hadamard product. Then, only the selected subset data in k-space is collected during the measurement. Thus, the accelerated image can be represented as:

$$\hat{x} = \mathcal{F}^{-1}(M \circ k) \tag{7}$$

This down-sampling process leads to less k-space data to collect and faster imaging speed. However, after applying the reverse Fourier transformation and go back to the image space, the resulting image typically includes aliasing artifacts. Heavier down-sampling typically leads to more intense aliasing artifacts and worse image quality. This k-space down-sampling can be considered as an image degradation.

Furthermore, consider a sampling mask $M_t$ that changes along time steps $t$. This image degradation can then be written as:

$$x_t = D(x_0, t) = \mathcal{F}^{-1}(M_t \circ k), \qquad t = 0, 1, \dots, T \tag{8}$$

In the cold diffusion model, the degradation severity varies along $t$. Where $t = 0$ corresponds to the original image and $t = T$ corresponds to the final degraded image. Here for the k-space cold diffusion, $x_0 = \mathcal{F}^{-1}(k)$ is the fully sampled image reconstruction and $t$ determines the down-sampling proportion in k-space. Larger $t$ number regarding a heavier down sampling in k-space. We set $M_{t=0} = J$, $M_{t=T} = M$, where $J$ denotes matrix of ones and $M$ denotes the sampling mask that acquires the measurement k-space data. This indicates that, for the un-degraded image at $t = 0$, the image is a fully sampled reconstruction from the k-space measurement, and the final degraded image is a zero-filled reconstruction from the sub-sampling measurement corresponding to the mask $M$. The sampling mask proportion for intermediate steps are scheduled linearly according to the step number $t$.

Later, we train a model to reverse this k-space degradation:

$$\hat{x}_0 = R(x_t, t) = R(\mathcal{F}^{-1}(M_t \circ k), t) \tag{9}$$

Then, we use equation (5) to predict the fully sampled reconstruction $\hat{x}_0$. Figure 1 illustrates the overall pipeline of our k-space cold diffusion model.

### 3. Implementation details

We performed our k-space down-sampling degradation with two types of masks: Cartesian sampling mask [29] and 2D Gaussian mask. Both are binary masks where 1 indicates that the corresponding k-space data is preserved and 0 indicates the corresponding k-space is masked out.

Let $M^c = \mathbf{I} - M$, which indicates the complement matrix of the mask $M$. At each time step, we randomly select a subset $M_t^c$ from $M^c$ such that the portion being selected is proportional to $(T-t)/T$, meaning $M_t^c$ is scheduled linearly for $t$. Then, $M_t = M + M_t^c$ and the corresponding image is $\mathrm{x}_t = \mathcal{F}^{-1}(M_t \mathrm{k}) = D(\mathrm{x}_0, t)$. Figure 2 further demonstrates examples for this k-space degradation process along time steps.

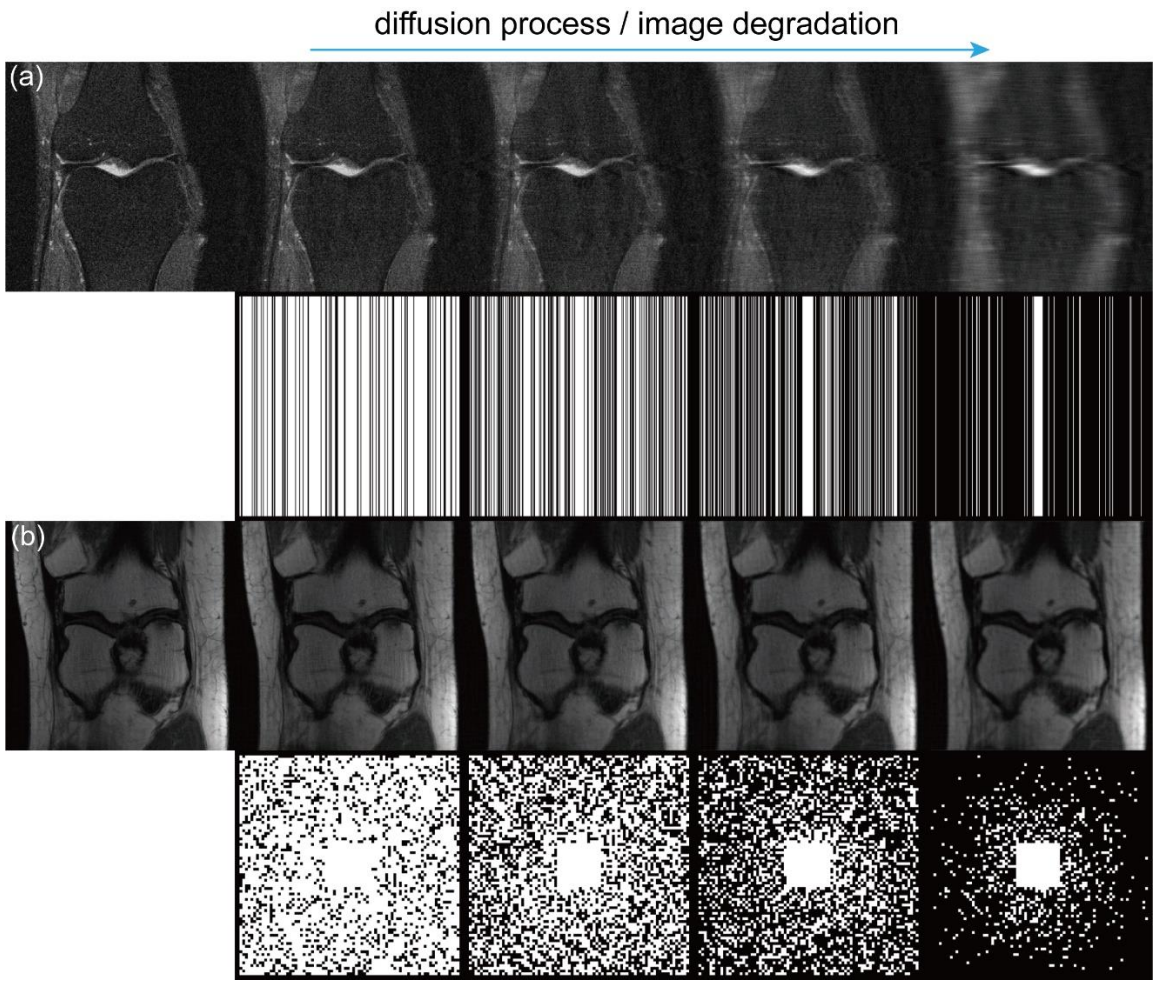

**Figure 2.** K-space cold diffusion degradation process. (a) for Cartesian sampling mask, (b) for Gaussian sampling mask.

We trained a U-Net [34] structure followed the cold diffusion model for image restoration. It includes 4 depth layers with the number of channels in the first layer being 64. The channel number doubles at each depth. We performed experiments on a large-scale open-source MRI dataset: fastMRI [29]. All experiments are performed with the single-coil knee dataset. Our k-space cold diffusion model was trained with k-space data computed from $320 \times 320$ complex images. For both acceleration masks, we performed 4-fold and 8-fold acceleration reconstruction, with the central fraction being 0.08 for 4-fold and 0.04 for 8-fold as suggested by the fastMRI work. We trained our k-space cold diffusion model by minimizing L1 loss in equation (3). The network was trained to predict the fully sampled reconstruction $\hat{x}_0$ given a k-space degraded image. All models were trained for 700000 iterations using the Adam optimizer [35]. The batch size is set to 6 and the learning rate is set to $2 \times 10^{-5}$.

To further demonstrate the effectiveness of our model, we performed comparison studies with multiple baseline deep learning-based MRI reconstruction models: U-Net, W-Net [5] and end-to-end variational net (E2E-VarNet) [36]. Note that although our model performs image degradation in k-space, the network works in the image space like the U-Net. Where W-Net and E2E-VarNet work in the k-space. All models have been set to have the same size for fair comparison. We report peak signal-to-noise ratio (PSNR) and structural similarity index (SSIM) [29] as our performance metrics since other baseline models utilize them for evaluation and comparison.

**Results**

All the experiments are performed with the large-scale open-source MRI dataset: fastMRI. We used the full single-coil knee dataset as our training set. For testing, we randomly selected 6 PD and PDFS volumes from the validation set.

Figure 3 demonstrates reconstruction results for 4-fold and 8-fold Cartesian under-sampling masks. We compare our k-space cold diffusion model with U-Net, W-Net and E2E-VarNet. Corresponding error maps are shown under each reconstruction image. All error maps have been magnified five times for better demonstration. We find that our k-space cold diffusion model preserves more image details and cleaner error maps compared to other models. Table 1 provides evaluation metrics for the 4-fold and 8-fold Cartesian under-sampling reconstructions, showing that our model outperforms others with a PSNR/SSIM of 30.58/0.7150 for 4-fold and 29.51/0.6414 for 8-fold. In Figure 4 and Table 2, we demonstrate reconstruction results and evaluation metrics for Gaussian sampling masks with 4-fold and 8-fold acceleration tasks. Our method outperforms others with a PSNR/SSIM of 30.31/0.7059 and 29.59/0.6416 for 4-fold and 8-fold, respectively. Similar to the Cartesian sampling reconstruction tasks, we found our model preserves more image details and textures and gives out cleaner error maps.

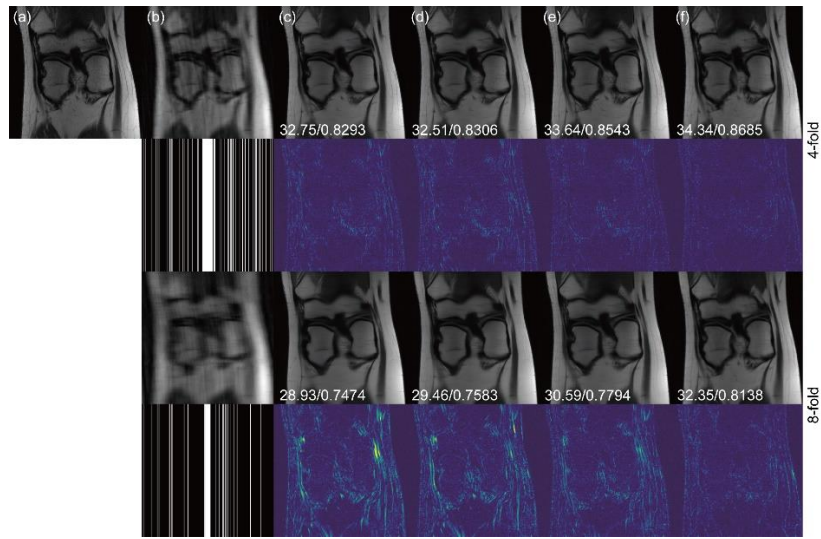

**Figure 3.** Reconstruction results for 4-fold and 8-fold accelerations with Cartesian sampling mask. (a) fully sampled target image, (b) zero-filled images and sampling masks, (c) U-Net reconstructions, (d) W-Net reconstructions, (e) E2E-VarNet reconstructions, (f) our proposed k-space cold diffusion reconstructions. Corresponding error maps are demonstrated together with their reconstructions. The numbers on the lower left of each image indicate PSNR and SSIM, respectively.

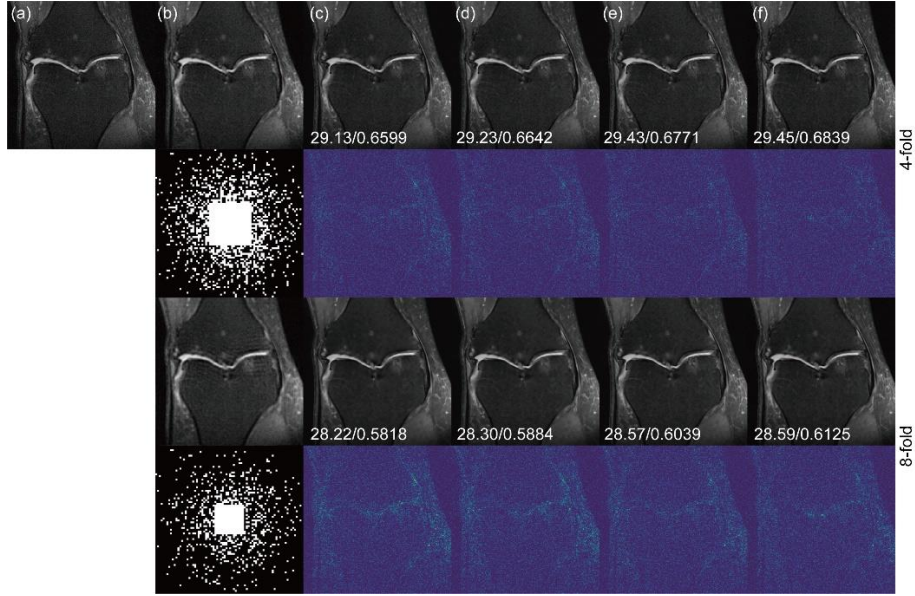

**Figure 4.** Reconstruction results for 4-fold and 8-fold accelerations with Gaussian sampling mask. (a) fully sampled target image, (b) zero-filled images and sampling masks, (c) U-Net reconstructions, (d) W-Net reconstructions, (e) E2E-VarNet reconstructions, (f) our proposed k-space cold diffusion reconstructions. Corresponding error maps are demonstrated together with their reconstructions. The numbers on the lower left of each image indicate PSNR and SSIM, respectively.

**Table 1.** Evaluation metrics for 4-fold and 8-fold 1D Cartesian under-sampling reconstructions.

|  | 4x | | 8x | |
| --- | --- | --- | --- | --- |
|  | PSNR | SSIM | PSNR | SSIM |
| U-Net | 28.21 | 0.6001 | 27.03 | 0.5400 |
| W-Net | 29.34 | 0.6464 | 28.42 | 0.6149 |
| E2E-VarNet | 30.29 | 0.6850 | 29.16 | 0.6338 |
| K-space cold diffusion | **30.58** | **0.7150** | **29.51** | **0.6414** |

**Table 2.** Evaluation metrics for 4-fold and 8-fold 2D Gaussian under-sampling reconstructions.

|  | 4x | | 8x | |
| --- | --- | --- | --- | --- |
|  | PSNR | SSIM | PSNR | SSIM |
| U-Net | 29.07 | 0.6244 | 28.24 | 0.5577 |
| W-Net | 29.76 | 0.6560 | 29.23 | 0.6071 |
| E2E-VarNet | 29.90 | 0.6621 | 29.42 | 0.6151 |
| K-space cold diffusion | **30.31** | **0.7059** | **29.59** | **0.6416** |

In order to explore the effects of the total sampling time steps on the final reconstructions, we performed tests using Cartesian sampling masks with the total time step $T$ set to 125, 250 and 1000. Corresponding evaluation metrics are shown in Table 3. We found that the PSNR and SSIM are rather close to each other in these settings. Note that for this Cartesian sampling case, the resolution for k-space is $320 \times 320$, meaning that for larger time steps, the same sampling proportion may map to the same time step, such that the final performance may stay the same. Yet for even higher resolution or heavier down-sampling ratio, larger time steps can still be beneficial. Here for our settings, we use $T = 125$ for all our tests as it takes less steps for reconstruction and is faster.

**Table 3.** Evaluation metrics for different time step settings. We performed tests using Cartesian sampling masks with the total time step $T$ set to 125, 250 and 1000.

|  | 4x | | 8x | |
| --- | --- | --- | --- | --- |
|  | PSNR | SSIM | PSNR | SSIM |
| T=125 | 30.58 | 0.7150 | 29.51 | 0.6414 |
| T=250 | 30.58 | 0.7148 | 29.50 | 0.6406 |
| T=1000 | 30.59 | 0.7149 | 29.51 | 0.6412 |

To further study the generalization capabilities of our k-space cold diffusion model, we performed zero-shot transfer learning tasks using equispaced Cartesian sampling, 1D Gaussian sampling, and 2D Gaussian sampling masks. In each setup, we utilized the models originally trained with Cartesian sampling masks, as demonstrated in Table 1. All models were evaluated directly with the corresponding sampling masks without any fine-tuning. Reconstruction comparison results are illustrated in Table 4. Our model achieves superior performance with a PSNR/SSIM of 30.54/0.7139 for 4-fold and 29.56/0.6440 for 8-fold equispaced Cartesian sampling masks. In the 1D Gaussian sampling zero-shot reconstruction test, our model outperforms others with a PSNR/SSIM of 30.34/0.7090 and 29.46/0.6464 for 4-fold and 8-fold, respectively. Similarly, in the 2D Gaussian sampling zero-shot test, our model maintains superior and stable performance with a PSNR/SSIM of 30.04/0.6992 for 4-fold and 29.31/0.6341 for 8-fold. Additionally, we conducted a 4-fold reconstruction test using 8-fold pre-trained models, as demonstrated in Table 4. Notably, our k-space cold diffusion model achieves a PSNR/SSIM of 30.50/0.7138 for equispaced Cartesian sampling, 30.17/0.7061 for 1D Gaussian sampling, and 30.04/0.6999 for 2D Gaussian sampling, which is comparable to the performance of the 4-fold model. Interestingly, the performance of other models significantly drops for 2D Gaussian sampling due to the difference in sampling strategies between 2D and 1D. This demonstrates that our model benefits from the k-space cold diffusion process, which is naturally and inherently conditioned on the sampling mask, allowing the model to be applied in a broader scope.

**Table 4.** Zero-shot learning evaluation metrics for equispaced Cartesian under-sampling, 1D Gaussian under-sampling, and 2D Gaussian under-sampling reconstructions.

| | Equispaced Cartesian Under-sampling | | | | | |
|---|---|---|---|---|---|---|
| | 4x | | 8x | | 4x with 8-fold model | |
| | PSNR | SSIM | PSNR | SSIM | PSNR | SSIM |
| U-Net | 28.56 | 0.6201 | 27.07 | 0.5406 | 27.53 | 0.6067 |
| W-Net | 29.63 | 0.6730 | 28.34 | 0.6109 | 28.56 | 0.6591 |
| E2E-VarNet | 30.25 | 0.6821 | 29.10 | 0.6210 | 29.53 | 0.6583 |
| K-space cold diffusion | **30.54** | **0.7139** | **29.56** | **0.6440** | **30.50** | **0.7138** |
| | 1D Gaussian Under-sampling | | | | | |
| | 4x | | 8x | | 4x with 8-fold model | |
| | PSNR | SSIM | PSNR | SSIM | PSNR | SSIM |
| U-Net | 28.70 | 0.6311 | 27.21 | 0.5579 | 27.53 | 0.6258 |
| W-Net | 29.61 | 0.6780 | 28.50 | 0.6253 | 28.32 | 0.6778 |
| E2E-VarNet | 30.19 | 0.6879 | 29.37 | 0.6309 | 29.47 | 0.6617 |
| K-space cold diffusion | **30.34** | **0.7090** | **29.46** | **0.6464** | **30.17** | **0.7061** |
| | 2D Gaussian Under-sampling | | | | | |
| | 4x | | 8x | | 4x with 8-fold model | |
| | PSNR | SSIM | PSNR | SSIM | PSNR | SSIM |
| U-Net | 27.90 | 0.6019 | 26.03 | 0.5193 | 26.28 | 0.5951 |
| W-Net | 28.97 | 0.6671 | 27.49 | 0.5989 | 27.45 | 0.6653 |
| E2E-VarNet | 28.30 | 0.6226 | 21.83 | 0.4067 | 20.82 | 0.4286 |
| K-space cold diffusion | **30.04** | **0.6992** | **29.31** | **0.6341** | **30.04** | **0.6999** |

As an image generation model, diffusion-based method can output multiple reconstructions given a starting point. The original noise-based diffusion models can generate multiple images from noises. Later conditional diffusion models are capable of performing multiple samplings given the same starting point. Here we also explore the effects of multiple samplings of our model. Examples are shown in Figure 5. For multiple sampling settings, we take the average of each pixel from different sampling images as the final output image. The uncertainty maps are calculated by taking the standard deviation among all samples in the pixel level. As the acceleration factor increases, the corresponding uncertainty increases as well. Corresponding PSNR/SSIM are labelled together with the image. We found that when the sampling number is 5-10, the improvements in metrics are rather noticeable. As the sampling number increases, the enhancement converges. Note that more samplings lead to a longer reconstruction time and the averaging over multiple samplings can lead to a smoothness for the final reconstruction, which hampers the image details. Taking these into consideration, we simply take single sampling during our reconstructions.

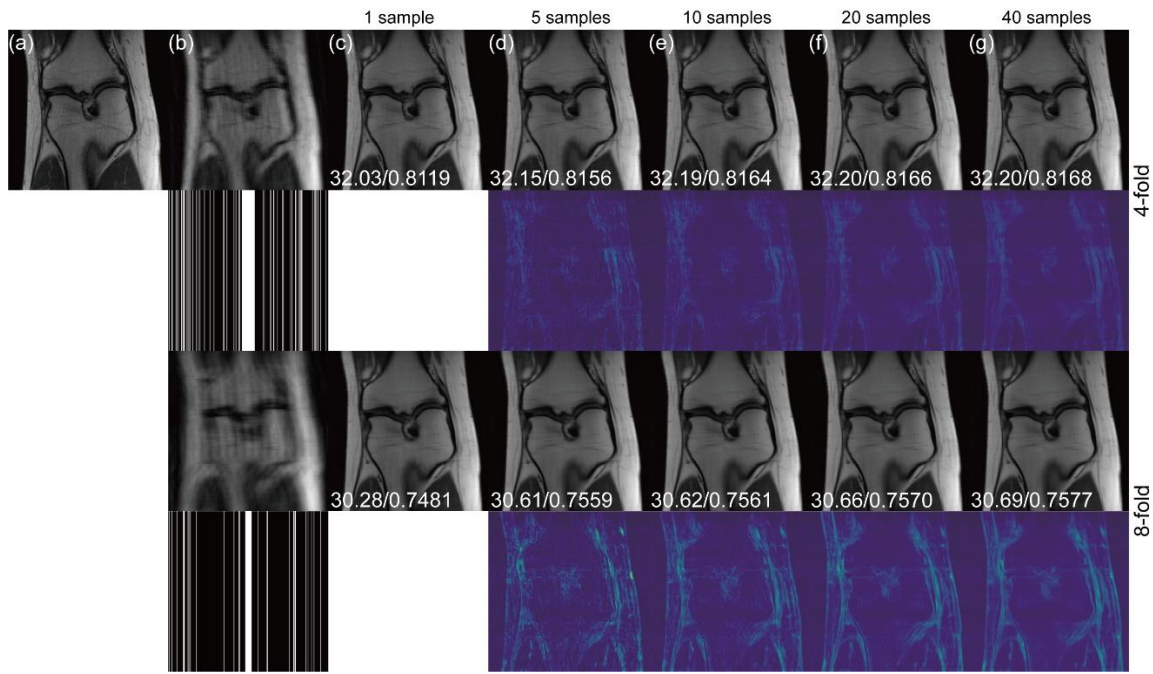

**Figure 5.** Reconstructions with different sampling numbers. (a) Fully sampled image, (b) zero-filled images and sampling masks. (c), (d), (e), (f) and (g) for 1, 5, 10, 20 and 40 samples for reconstructions. Images are generated by taking the average of multiple samples. Corresponding uncertainty maps are shown under each image. The numbers on the lower left of each image indicate PSNR and SSIM, respectively.

**Conclusion**

Under-sampling reconstruction is an essential topic for imaging. Diffusion-based models have shown superior performance in multiple fields nowadays including image generation, super-resolution, in-painting, editing, text-to-image transformation, etc. Moreover, cold diffusion provides a framework for generalized diffusion models. The image degradation is no longer limited to Gaussian noise, but other general image degradation operations such as blurring, inpainting, down-scaling, and snowification, among others. In this work, we present a novel k-space cold diffusion framework for accelerated MRI reconstruction. Instead of performing image degradation in image space, k-space cold diffusion performs image degradation by continuously increasing the down-sampling ratio in k-space and using a deep neural network to learn the restoration. This k-space degradation can accommodate different sampling masks. Furthermore, by embedding the k-space sampling process directly into the image degradation process, k-space cold diffusion is inherently conditioned on the sampling mask. A deep neural network is trained to effectively reverse this process, thereby enhancing the model's ability to generalize. Although we present our work with the MRI acceleration task, we want to note that this method can applied to other medical imaging fields such as the sparse view CT reconstruction.

We tested our k-space cold diffusion method with a large-scale open-source dataset with different acceleration factors and sampling masks. Our results show that our method outperforms other typical reconstruction networks working in image space, k-space and unrolled structure. We conducted generalizability tests for zero-shot transfer learning using unforeseen masks. Remarkably, our model demonstrated superior performance and maintained similar levels of accuracy in 4-fold reconstructions, even when using 8-fold pre-trained models. Furthermore, we explore the effects of the total time steps and sampling number and show that the method is capable of generating good results even with a small total time step and just a single sample.

There remains much potential for future works. For example, the current k-space cold diffusion model uses linear scheduling for k-space degradation following the leads in the original cold diffusion work. However, lower frequency regions in k-space correspond to the overall image content in the image space, while higher frequency regions correspond to image details. Thus, the degradation process can be further accommodated to encounter this difference. Secondly, the k-space degradation process can be further pixelized to enable a more refined restoration process. This can be especially beneficial if the acceleration factor is large, and the sampling mask is unknown. Another interesting topic can be a latent space cold diffusion work for accelerated image reconstruction. This would further improve the efficiency of image restoration and can enable prompt design for more general reconstruction tasks.

**Acknowledgements**

This work was supported by the Rajen Kilachand Fund for Integrated Life Science and Engineering. We would like to thank the Boston University Photonics Center for technical support.

**Author contributions**

Conceptualization: G.S., S.A., X.Z.; Methodology: G.S., M.L., S.A.; Software: G.S., M.L.; Formal Analysis: G.S., C.W.F.; Writing – Original Draft: G.S., M.L.; Writing – Review & Editing: C.W.F., S.A., X.Z.; Project Administration: X.Z.; Funding Acquisition: X.Z.

**Competing interests**

The authors declare no competing interests.

**Data availability statement**

The fastMRI dataset that support the findings of this study are openly available at: https://fastmri.med.nyu.edu/. The codes of this study are openly available at: https://github.com/GuoyaoShen/K-SapceColdDIffusion.